\documentclass[10pt, aps, prd, amssymb, amsmath, amsfonts, showpacs, floatfix, twocolumn, twoside, a4paper, superscriptaddress, longbibliography, nofootinbib]{revtex4-2}

\usepackage{graphicx}
\usepackage[usenames,dvipsnames]{xcolor} 
\usepackage[utf8]{inputenc}
\usepackage{color}
\usepackage{hyperref}
\usepackage[normalem]{ulem} 
\usepackage{physics}
\usepackage[capitalise]{cleveref}
\usepackage{mathrsfs, mathtools}
\usepackage{siunitx}
\usepackage{comment}
\usepackage{soul}
\usepackage[T3,T1]{fontenc}
\DeclareMathAlphabet{\mathpzc}{OT1}{pzc}{m}{it} 
\usepackage{mathrsfs} 
\usepackage{bm} 
\usepackage{microtype} 
\def\MT@register@subst@font{
	\MT@exp@one@n\MT@in@clist\font@name\MT@font@list
	\ifMT@inlist@\else\xdef\MT@font@list{\MT@font@list\font@name,}\fi}
\makeatother
\usepackage{orcidlink}
\usepackage{multirow}
\usepackage{booktabs} 
\usepackage{array} 
\usepackage{tabularx}
\usepackage{braket} 
\usepackage{dblfloatfix}

\newcolumntype{C}{>{\centering\arraybackslash}X}

\usepackage{hyperref} 
\usepackage{nicefrac} 

\usepackage{enumitem}
\usepackage{comment}
\usepackage{comment}
\usepackage{xspace}
\usepackage{bm}
\hypersetup{
  colorlinks   = true, 
  urlcolor     = blue, 
  linkcolor    = blue, 
  citecolor   = blue 
}
\allowdisplaybreaks[1]
\graphicspath{{figures/}}

\usepackage{float}

\makeatletter
\def\@setemails{\vskip-20pt\@emails} 
\makeatother

\begin{document}

\title{Black hole binaries in shift-symmetric Einstein-scalar-Gauss-Bonnet gravity}

\author{Maxence Corman}
\email{maxence.corman@aei.mpg.de}
\affiliation{Max Planck Institute for Gravitational Physics (Albert Einstein Institute), D-14476 Potsdam, Germany}

\author{Llibert Arest\'e Sal\'o}
\email{llibert.areste@uib.cat}
\affiliation{Institute of Applied Computing \& Community Code (IAC3), Universitat de les Illes Balears, Palma de Mallorca, E-07122, Spain}
\affiliation{Instituut voor Theoretische Fysica, KU Leuven. Celestijnenlaan 200D, B-3001 Leuven, Belgium. }
\affiliation{Leuven Gravity Institute, KU Leuven. Celestijnenlaan 200D, B-3001 Leuven, Belgium. }

\author{Katy Clough}
\email{k.clough@qmul.ac.uk}
\affiliation{School of Mathematical Sciences, Queen Mary University of London, Mile End Road, London, E1 4NS, United Kingdom}

\begin{abstract}
In shift-symmetric Einstein-scalar-Gauss-Bonnet gravity, stationary black holes have a non-vanishing scalar charge. 
During the inspiral, the phase evolution is modified by several effects, primarily an additional scalar dipole radiation, which enters at -1PN order. Including corrections up to 2PN, this effect accelerates the inspiral when compared to general relativity, but changes in the conservative dynamics may do the opposite.
Using fully non-linear numerical simulations of quasi-circular, comparable mass binaries, we characterize the late stages of the orbital dynamics. We find that the overall effect is still an accelerated merger phase for the modified gravity case, but noticeably less than predicted by PN alone, which we relate to the conservative dynamics, showing that at the late inspiral stage more energy must be emitted in scalar-Gauss-Bonnet gravity to induce a given change in frequency.
In longer signals, this may lead to a distinctive frequency evolution relative to general relativity as the binary approaches merger, but its effect close to merger is to suppress the difference with GR. This work suggests we may need to revisit existing constraints on the theory that are obtained assuming PN approximations apply up to merger, or based on order by order approximations that neglect backreaction effects on the metric, and shows the importance of including non-linear effects that modify the gravitational sector in the strong field regime.
\end{abstract}

\maketitle

\section{Introduction}

Gravitational waves (GWs) from compact binary coalescences provide a unique probe of the strong field regime of black holes (BHs) \cite{LISA:2022kgy,Gnocchi:2019jzp,Barack:2018yly}. Observations by the LIGO-Virgo-KAGRA network of detectors are currently ongoing \cite{LIGOScientific:2025snk,LIGOScientific:2025bkz}, and the future proposed detectors LISA \cite{LISA:2017pwj}, the Einstein Telescope \cite{Punturo:2010zz,ET:2019dnz}, and Cosmic Explorer \cite{Reitze:2019iox} will provide a significant upgrade in the precision that can be achieved. These messengers carry information about a previously unexplored regime of parameter space, probing stronger potentials and higher curvatures than previous observations \cite{Baker:2014zba}, and therefore, there is a possibility that new effects from high-energy corrections to general relativity (GR) may be observed. 

Based on the data we have to date, any deviations are expected to be small \cite{LIGOScientific:2019fpa,Abbott:2020jks,LIGOScientific:2020tif,LIGOScientific:2021sio, LIGOScientific:2025obp,LIGOScientific:2025rid}. Effective Field Theories (EFTs) of gravity therefore provide a well-motivated way to organize and constrain corrections to GR. A given EFT specifies any lower energy degrees of freedom and symmetries that should be respected, and then parametrizes the leading order terms in the action that modify GR in powers of a dimensionful coupling $\lambda$ that is small at the relevant curvature scale (e.g. for a BH, this scale is related to the inverse of the Schwarzschild radius). Constraints on the coupling can then be translated into bounds on possible UV completions and their properties (see e.g. \cite{deRham:2017avq,deRham:2021bll,CarrilloGonzalez:2022fwg,CarrilloGonzalez:2023cbf}).

Einstein-scalar-Gauss-Bonnet (EsGB) gravity considers the effect of an additional scalar degree of freedom, which couples to the Gauss-Bonnet invariant via an arbitrary coupling function which can be expressed from an EFT point of view as a power expansion of the scalar. In the shift-symmetric case, this function is linear, which represents the leading order parity-invariant correction to GR for a scalar-tensor theory of gravity in vacuum \cite{Weinberg:2008hq}. In such a theory, the scalar is sourced by spacetime curvature and all BHs should carry a scalar charge \cite{Kanti:1995vq,Sotiriou:2013qea,Sotiriou:2014pfa}. As a result, the theory is relatively well constrained by observations, but tests so far rely mainly on Post-Newtonian (PN) approximations \cite{Lyu:2022gdr,Julie:2024fwy,Sanger:2024axs} that cannot capture the fully non-linear regime near merger, or order-reduced approximations that are known to suffer from spurious secular growth \cite{Reall:2021ebq}. Consequently, there has been a significant effort to develop fully non-linear simulations of binaries in this and other beyond-GR theories using numerical relativity (NR), in order to generate more accurate waveforms.

We would like to know if modifications to GR are detectable, which is not the same as seeing that the waveform differs from its GR equivalent with all other parameters held equal. The real test is whether the corrected waveforms are distinguishable from GR signals when injected into the data-stream. It is possible that, whilst they may differ, the beyond-GR effects simply change the inference of the intrinsic binary parameters. If the residual, once the best fit GR waveform is subtracted, is small, it may not be significant enough to trigger a detection with current detector sensitivities. This could be the case where, for example, there is a simple overall increase in the rate of dephasing, which could be mimicked by larger BH masses.

A major breakthrough enabling the construction of injectable signals was the demonstration by Kovács and Reall that some higher-derivative scalar–tensor theories admit a well-posed initial value formulation in the weakly coupled regime when expressed in modified generalized harmonic (MGH) coordinates \cite{Kovacs:2020pns,Kovacs:2020ywu}. This formulation permitted the first fully non-linear evolutions in such theories \cite{East:2020hgw} and, together with variations for puncture based codes using a modified CCZ4 formalism (mCCZ4)~\cite{AresteSalo:2022hua,AresteSalo:2023mmd} and a modified BSSN formalism (mBSSN)~\cite{Shum:2025lgp}, this has enabled further numerical work \cite{East:2021bqk,East:2022rqi,Corman:2022xqg,Doneva:2023oww,Corman:2024vlk,Doneva:2024ntw,AresteSalo:2025sxc,Capuano:2026lhs,AresteSalo:2026zzc,Swain:2026dfr} and the first quantification of error in alternative methods \cite{Corman:2024cdr}, such as the order by order approach \cite{Okounkova:2020rqw,Okounkova:2017yby,Silva:2020omi,Elley:2022ept,Doneva:2022byd,Evstafyeva:2022rve} and the hydrodynamics inspired ``fixing the equations'' technique \cite{Bezares:2021yek,Cayuso:2023xbc,Cayuso:2020lca,Cayuso:2023dei,Franchini:2022ukz}. Initial simulations mainly focused on the tendency of evolutions to remain in the weakly coupled regime in which the EFT is valid in dynamical strong field cases \cite{East:2021bqk,R:2022hlf,Doneva:2023oww} and proof of principle studies of binary evolutions \cite{East:2020hgw,East:2021bqk,East:2022rqi,AresteSalo:2023mmd}. More recent works have attempted to render the simulations more precise, but have encountered difficulties in controlling the initial data and eccentricity, as well as quantifying the small effects that arise due to numerical errors \cite{Corman:2022xqg,AresteSalo:2025sxc}. Improvements to the techniques that would enhance the quality of initial data are being developed \cite{Brady:2023dgu,Nee:2024bur}, but the problem remains challenging.

\begin{figure}[t]
\includegraphics[width=\columnwidth]{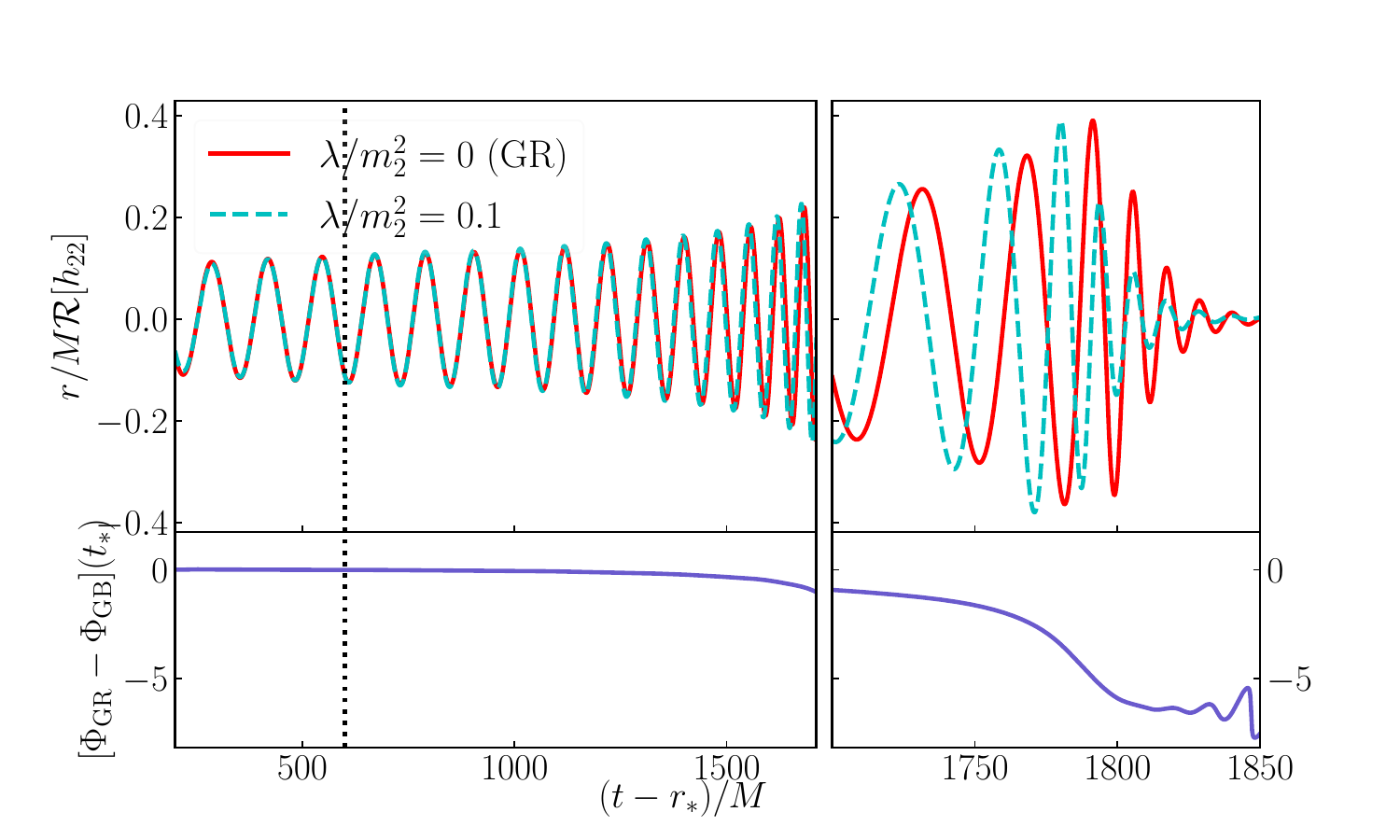}
\caption{Here we show the strain waveform for EsGB and GR for the intrinsic parameters of a GW150914-like event using the mCCZ4 formulation. The signals have been aligned in frequency and phase according to \eqref{eq:align_freq_1}-\eqref{eq:align_freq_2} over time window ranging from $t_i=200M$ to $t_f=600M$, the latter shown by the vertical dotted line. One can see that the GR waveform merges \textit{later} than the EsGB case, but the difference is small. The bottom panel shows the dephasing as a function of retarded time $t_*\equiv(t-r_*)/M$. 
}\label{fig:Strain}
\end{figure}

In this article we present calibrated simulations for a binary with intrinsic parameters similar to the first detected GW event, GW150914 \cite{LIGOScientific:2016aoc}, and compare the GR and EsGB waveforms. These results indicate that the modified theory exhibits an overall acceleration in the phase evolution during the late inspiral phase, as shown in Fig.~\ref{fig:Strain}. We compare this to PN calculations and find that it is noticeably lower than the prediction at 2PN. We verify this result using the two independent codes that have implemented the MGH and mCCZ4 formulations of the equations, and identify the reduction in the dephasing compared to PN with a change in the conservative dynamics, where more energy must be emitted in the modified theory to induce a given change in the frequency. Although our waveforms are too short to follow this, we expect the early inspiral to be dominated by the dissipative effects of the scalar radiation, and thus for the phase evolution to be similar to the PN prediction for the same intrinsic parameters. This change in dephasing behaviour from the PN predictions from inspiral to merger may then be distinctive from an observational perspective.

We use geometric units: $G=c=1$, the metric convention $-+++$, 
and lower case Latin letters to index
spacetime indices. The Riemann tensor is $R^a{}_{bcd}=\partial_c\Gamma^a_{db}-\cdots$.

\section{EsGB theory}
We consider shift-symmetric EsGB gravity, which has an action given by
\begin{align}
\label{eq:esgb_action}
    S
    =&
    \frac{1}{16\pi}\int d^4x\sqrt{-g}
   \left(
        R
    -   \left(\nabla\phi\right)^2
    +   2\lambda\,\phi\,\mathcal{G}
    \right)
    ,
\end{align}
where $g$ is the determinant of the spacetime metric and
$\mathcal{G}$ is the Gauss-Bonnet scalar given by $
    \mathcal{G}
    \equiv
    R^2 - 4R_{ab}R^{ab} + R_{abcd}R^{abcd}
    .$
Here, $\lambda$ is a positive constant coupling parameter that, in
geometric units, has dimensions of length squared\footnote{The coupling function to the Gauss-Bonnet scalar can be expressed from an EFT point of view as a power expansion of the scalar and, thus, the leading order effect is expected to be seen from the linear term.}.
As the Gauss-Bonnet scalar $\mathcal{G}$ is a total derivative
in four dimensions, we see that the action
is preserved
up to total derivatives under constant shifts in the scalar field:
$\phi\to\phi+\textrm{constant}$.
Schwarzschild and Kerr BHs are not stationary solutions
in this theory: if one begins with such vacuum initial data, the BHs will
dynamically develop stable scalar clouds (hair).
The end state is a BH with non-zero scalar charge $Q_{\rm SF}$,
such that at large radius the scalar field falls off like
$\phi = \phi_0 + Q_{\rm SF}/r + \mathcal{O}(1/r^2)$, $\phi_0$ being the scalar field value at spatial infinity imposed by the binary's cosmological environment.
Studies have found that stationary solutions exist and have computed properties of the black hole horizon and surrounding scalar charge,
as long as the coupling normalized by the total BH mass as measured
at infinity $m$,
$\lambda/m^2$, is sufficiently small
\cite{Sotiriou:2013qea,Sotiriou:2014pfa,Ripley:2019aqj,East:2020hgw,Ripley:2019irj,Julie:2022huo}.
One expects BH binaries to emit scalar radiation, thus increasing the inspiral rate of binaries \cite{Yagi:2015oca,Yagi:2011xp}.
The most stringent observational bound on the theory, assuming the background scalar field vanishes, comes from comparing the black hole-neutron star GW signal GW230529 to Post-Newtonian results for EsGB and give a constraint of (after restoring dimensions) $\sqrt{\lambda}\lesssim 0.63$ km \cite{Sanger:2024axs}.  Note we assume $\lambda$ is positive without loss of generality. The sign determines the sign of the scalar profile that is dynamically generated, and so if $\lambda$ takes the opposite sign, so will the scalar field $\phi$ and the two signs cancel.  The coupling constant has typically been assumed to be positive in the literature, consistent with Einstein-dilaton-Gauss-Bonnet being a low-energy limit of string theory.
The equations of motion for shift-symmetric EsGB gravity are
\begin{align}
\label{eq:eom_esgb_scalar}
   &\Box\phi
   +
   \lambda\mathcal{G}
   =
   0
   ,\\
\label{eq:eom_esgb_tensor}
   &R_{ab}
   -
   \frac{1}{2}g_{ab}R
   -
   \nabla_a\phi\nabla_b\phi
   +
   \frac{1}{2}\left(\nabla\phi\right)^2g_{ab}
   + \nonumber \\
   & \quad \quad \quad \quad \quad \quad \quad \quad 2\lambda
    \delta^{efcd}_{ijg(a}g_{b)d}R^{ij}{}_{ef}
   \nabla^g\nabla_c\phi
   =
   0
   ,
\end{align}
where $\delta^{abcd}_{efgh}$ is the generalized Kronecker delta tensor.

\section{Formulation and methods}
To study the full, non-perturbative, shift-symmetric EsGB equations, we employ two independent formulations of the Einstein equations, each with its own numerical implementation: the MGH formulation~\cite{Kovacs:2020pns,Kovacs:2020ywu} used in \cite{East:2020hgw}, and the modified CCZ4 (mCCZ4) formulation~\cite{AresteSalo:2022hua,AresteSalo:2023mmd} implemented in \texttt{GRFolres}~\cite{AresteSalo:2023hcp} as an extension to \texttt{GRChombo}~\cite{Andrade:2021rbd,Radia:2021smk}.
Both formulations introduce two auxiliary
Lorentzian metrics to fix the gauge, 
high-order finite differencing with Runge–Kutta time integration, and adaptive mesh refinement, but they differ in important aspects.
The MGH scheme evolves the covariant system directly in compactified coordinates with a modified version of the damped harmonic gauge and relies on dynamic excision of regions within apparent horizons, whereas the mCCZ4 approach uses a 3+1 conformal decomposition on a finite Cartesian grid with moving-puncture gauge conditions. A detailed comparison of the formulations and numerical methods is provided in the Appendices.

For initial data, we start with a trivial $\phi$ profile, choosing
$\phi=\partial_t \phi = 0$, in which case the constraint equations of EsGB are the
same as in GR (see Ref.~\cite{East:2020hgw}). 
We can then start with constraint satisfying binary vacuum GR BH
solutions, constructed using the BH
puncture method with the \texttt{TwoPunctures}
\cite{Ansorg:2004ds,Paschalidis:2013oya,code_repo_tp} code.
For the scalar field, we slowly ramp up the coupling of the theory as described
in Appendix B of Ref.~\cite{Okounkova:2019zjf} over a period of time (typically $\sim 150M$, with $M$  the total ADM mass of the spacetime).
We study a system with parameters consistent with GW150914 \cite{LIGOScientific:2016vlm},
specifically, those used in Fig.~1 of the GW150914 detection paper \cite{LIGOScientific:2016aoc}. We set the spins to zero and the mass ratio to $q=1.221$, giving initial puncture mass values of $m_1= 0.5497M$ and $m_2 = 0.4502M$.
We choose an initial separation of $11.8M$ such that, in GR, the BHs merge
after $\sim 9$ orbits (roughly $1800M$). We consider a coupling value of
$\lambda/m_2^2=0.1$, which, assuming the smallest BH observed so far is $3.6 M_{\odot}$, corresponds to $\sim 1.7$ km, i.e. larger by a factor of $\sim 2$ than the current bounds we have on this theory. 
While such a value is ruled out, using a larger value helps us to see better the differences with GR, which, as noted, are smaller than expected.
We estimate the initial orbital eccentricity of our GR runs to be
$\sim 10^{-3}$. Without any adjustment of the parameters, the $\lambda/m_2^2=0.1$ runs result in initial orbital eccentricities of $\sim 10^{-2}$ due to being
perturbed by the development of the scalar field around the BHs at early times, which is an artefact of the initial conditions we use. In previous studies \cite{Corman:2022xqg,Corman:2024cdr,Corman:2024vlk,AresteSalo:2025sxc} it was found that residual eccentricity from the initial data was sub-dominant to finite-resolution truncation errors, because the energy contained in the scalar cloud, for the couplings and initial separations considered, was only a small fraction of the total binary binding energy. However this is no longer true for the larger initial separation and coupling values we consider in this work. Therefore, we perform eccentricity reduction following the methods of~\cite{Buonanno:2010yk,Habib:2024soh} and reduce the eccentricities to $\sim 2 \times 10^{-3}$ and $\sim 3 \times 10^{-3}$ for the MGH and mCCZ4 codes respectively.

We use many of the same diagnostics introduced in Refs.~\cite{East:2020hgw,Corman:2022xqg}.
We measure the scalar and gravitational radiation by extracting the Newman-Penrose scalar $\Psi_4$ and scalar field $\phi$ on coordinate spheres at large radii (typically $r = 100M$) and decomposing these quantities into spin $-2$ and $0$ weighted spherical harmonics. We use $\Psi_4$ to calculate an associated gravitational wave luminosity $P_{\rm GW}$ and $\phi$ to measure the flux of energy in the scalar field $P_{\rm SF}$ (cf. Eq.~(6) and Eq.~(9) in \cite{Corman:2022xqg}).
We neglect effects due to slightly different gauges/physical extraction radii at this
time~\cite{Lehner:2007ip}; at such distances, their effect
would be significantly smaller than the measured physical waveforms. We also verified that extrapolating our waveforms at infinity does not affect the phase and only slightly changes the amplitude, so considering finite radii is sufficient for the purposes of this work.

We calculate the gravitational wave strain $h \equiv h_{+} + i h_{\times}$, related to $\Psi_4$ through $\Psi_4 = \ddot{h}$ using fixed frequency integration \cite{Reisswig:2010di}. We further decompose the $h_{22}(t)$ mode into a real time-domain amplitude
$A(t)$ and phase $\Phi(t)$ as
$h_{22}(t) = A(t) e^{-i \Phi(t)}$
giving a GW frequency (assuming the stationary phase approximation) $\omega (t) =  2\pi f(t) = \frac{d}{dt}\Phi(t)$. 
During the evolution, we track any apparent horizons (AH) present at a given time, and measure their corresponding areas $\mathcal{A}_{\rm H}$.
From this, we compute a measure of the BH mass $M_{\rm H}$ via the Christodoulou formula \cite{Christodoulou:1970wf}. We provide further comments on this measure in the next section.

\section{Definition of dephasing and mass in different theories}
Taking the view of an observer far from the binary measuring the signal, we see a wave with an initial frequency $f_1$ and follow it up to a frequency $f_2$ (or merger). Say we observe $n$ cycles in GR. If EsGB modifies the binary evolution, we expect to measure a different number of cycles in this frequency interval. This tells us that the rate of inspiral has changed. But to compare the two systems meaningfully in this way, we should fix the intrinsic binary parameters to be the same in both cases. Our finding therefore rests in part on the definition of ``like-for-like'' intrinsic parameters in GR versus EsGB.
Since the systems are physically different and evolve dynamically in different ways, it is not clear that there is a unique way to compare them. However, having fixed the spins to be zero and reduced the initial eccentricities to be roughly zero, the problem reduces to the question of what we mean by equivalent masses. From an NR perspective the only gauge invariant notion of the mass of the spacetime is given by the total ADM mass $M_{\rm ADM}$, so in our simulations we fix this to be the same in both cases in the initial data. This also fixes the units of time of the asymptotic observer -- in the asymptotic region, the scalar field value is negligible and we are essentially in GR - here Newton's constant $G=1$ controls the mapping between energy/mass and time/length, so units are the same in GR and EsGB for a fixed $M_{\rm ADM}$. However, note that in the EsGB case this means that after the initial transient scalarization, some of the mass energy will be in the scalar charge, and the quasi-local masses associated with the horizon areas of the BHs via the Christodoulou formula will be different. The presence of a scalar charge also means that a given observed frequency will correspond to a different orbital separation (loosely speaking, since separation is gauge dependent in the strong field regime), as the local effective strength of gravity is modified. We will discuss this further in Sec. \ref{sec:pn_mass} once we have presented the results.

\vskip 5pt
\section{Results}
Our main result is shown in Fig.~\ref{fig:Strain}, where we compare the strains for the GR and EsGB cases as a function of time after aligning them in frequency and phase over an initial time interval, and observe an earlier merger in the EsGB case. We see that the GR binary is spiralling in more slowly - $df/dt = d^2\Phi/dt^2$ is \textit{smaller}, such that after the alignment (which sets the initial phase and its first time derivative equal) $\Phi_{\rm GR}(t_*) > \Phi_{\rm GB}(t_*)$ at any given retarded time, $t_* \equiv (t-r_*)/M$. 
In the Appendices we provide further details on the comparison and validation of our codes, including robustness studies for the alignment method used.

\begin{figure}[t]
\includegraphics[width=\columnwidth]{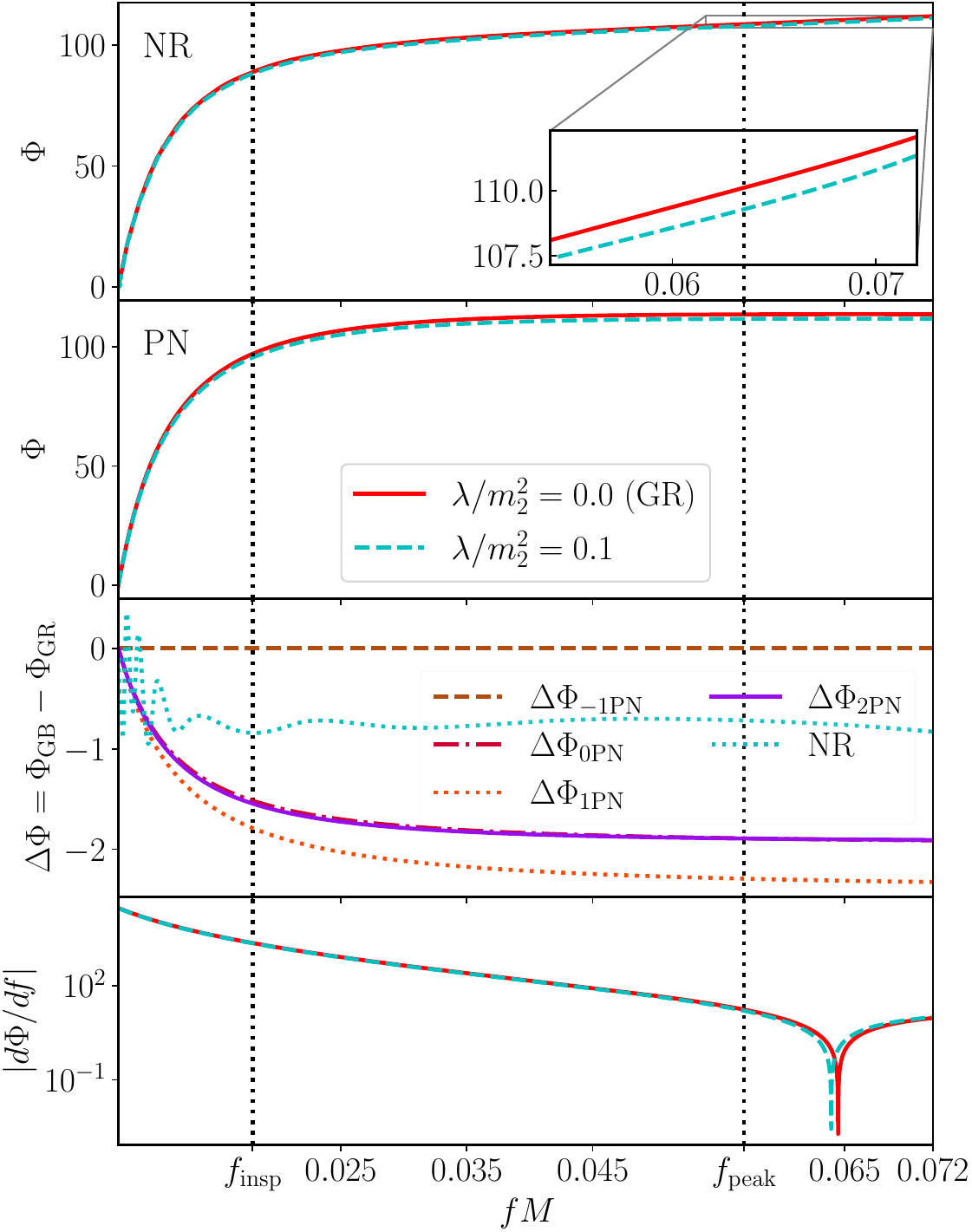}
\caption{Comparison of our NR results to PN predictions for the phase in EsGB and GR using same intrinsic parameters and range of frequencies probed by system studied in this paper. 
Top: Phase as a function of frequency in both GR (solid) and EsGB (dashed) gravity for the NR simulations.  
Top middle: The same but for the PN predictions, showing the larger difference.
Bottom middle: Net dephasing in the NR simulations and at different PN orders (the dephasing contains leading order (dipolar) contributions to phase and all contributions up to the given orders). 
Bottom: The PN prediction for $d\Phi/df \sim (df/dt)^{-1}$ as a function of frequency for GR and EsGB. The first vertical line corresponds to the end of inspiral stage in GR $f_{\rm insp}=0.018M$, and the second to frequency at which amplitude of GR waveform peaks $f_{\rm peak}$. The smaller value for EsGB implies a faster inspiral.
}\label{fig:PN}
\end{figure}

In Fig.~\ref{fig:PN} we compare to PN theory predictions for the dephasing for the system we study. The gravitational modes $h_{\ell m}$ have been computed to 2PN (relative to the quadrupolar radiation in GR) and we refer to Ref.~\cite{Julie:2024fwy} for complete expressions or Section III.D of Ref.~\cite{Corman:2024vlk} for a summary of the PN predictions. Substituting the intrinsic parameters of the system we study, the top panel of Fig.~\ref{fig:PN} shows the measured phase evolution as a function of frequency from our NR simulations, and the second shows the PN prediction up to 2PN all the way up to merger, in which the difference is visibly larger. Note that at any given frequency $\Phi_{\rm GB}(f)>\Phi_{\rm GR}(f)$, meaning that the EsGB waveform has to go through fewer cycles than in GR to reach a given frequency, which is in agreement with our finding\footnote{Note that at a fixed frequency $d\Phi/df \sim (df/dt)^{-1}$, which explains the opposite sign of the phase difference compared to the time dependence shown in Fig.~\ref{fig:Strain}.}, but the magnitude of dephasing we observe is clearly significantly smaller (the NR line in the third plot is less than half the 2PN prediction). Note that in the effective-one-body (EOB) treatment of Ref.~\cite{Julie:2024fwy}, the predicted GR--EsGB dephasing is much smaller than that of the PN series. We leave a comparison of our waveforms with EOB for future work.

The PN approximations predict that the EsGB system should inspiral faster than GR at any given frequency all the way up to near merger, but this effect diminishes at high frequencies, and the highest PN corrections, namely 1.5 and 2PN, reduce the rate of inspiral with respect to GR, which was already observed in Ref.~\cite{Corman:2024vlk}, as it is here. In the last panel we show the PN predictions for the quantity $d\Phi/df \sim (df/dt)^{-1}$ at any given frequency, which is larger in GR, again indicating that the EsGB waveform inspirals faster when compared to GR.

\begin{figure}[t]
 \includegraphics[width=0.99\columnwidth]{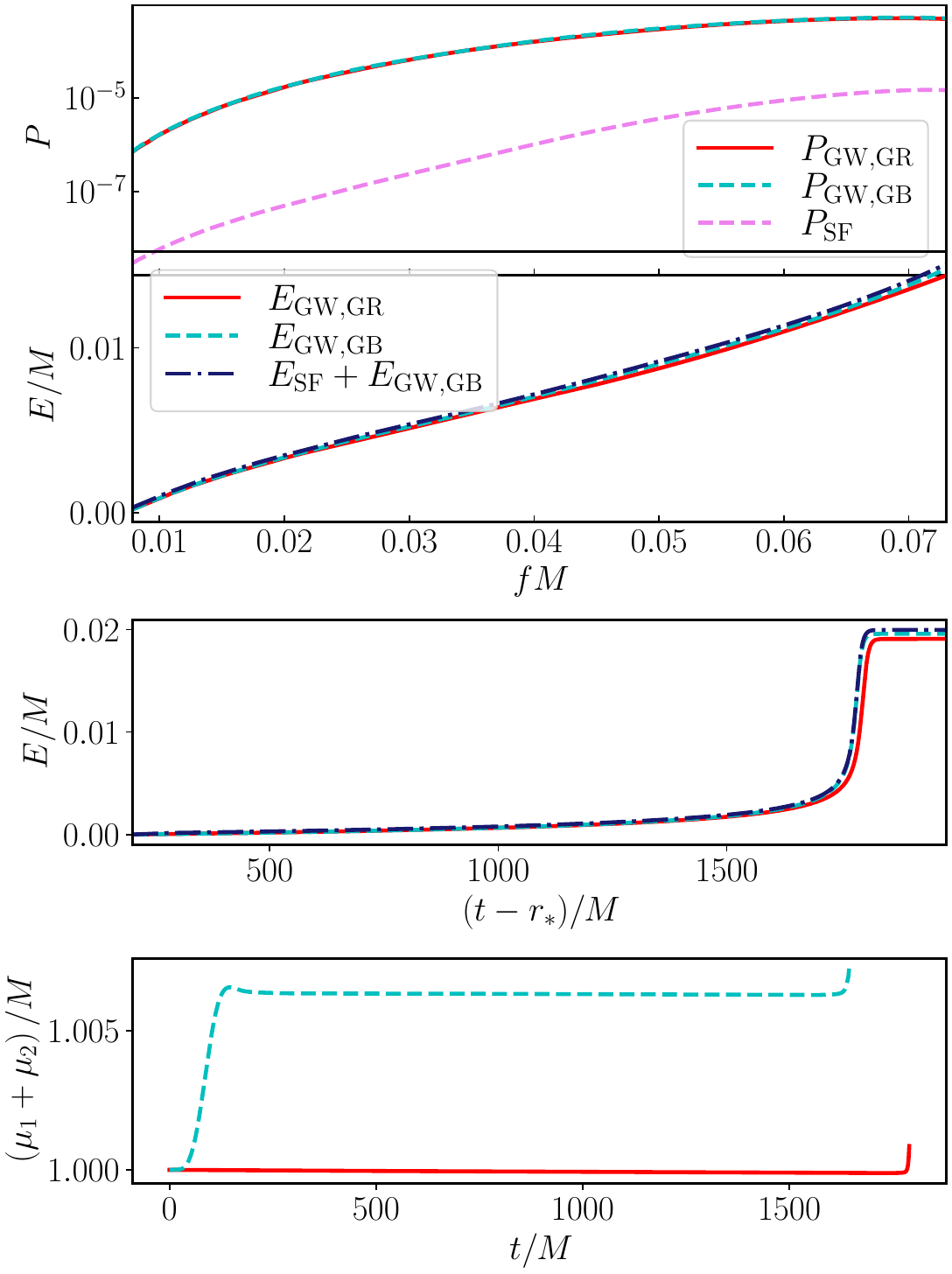}
\caption{Measured related to the conservative dynamics of the merger. 
Top: The power emitted in scalar and GWs in the GR and EsGB simulations. 
Top middle: The integrated energy emitted in these channels.
Bottom middle: The integrated GW luminosity versus retarded time - we see that the total energy emitted in the GR case is lower.
Bottom: The irreducible masses of the BHs in each case, showing the increase in this measure of the mass for EsGB.
}\label{fig:GWPower}
\end{figure}


To understand the origin of the dephasing better, in Fig.~\ref{fig:GWPower}, we consider the evolution of several quantities with respect to frequency and time. In the top panel we plot the GW luminosity $P_{\rm GW} = dE_{\rm GW}/dt$ as a function of frequency $f$, which is found to be similar between GR and EsGB. The flux of energy in the scalar field, $P_{\rm SF}$, is subdominant by two orders of magnitude, meaning we are in the so-called \textit{quadrupolar driven regime} \cite{Sennett:2016klh}. 
In general the quadrupole radiation should depend non-trivially on the values and gradients of the scalar at the late inspiral/merger stage, and it is therefore not obvious that the power as a function of frequency should remain unchanged, but it appears to be the case here. Additional energy emission from the scalar tends to accelerate the EsGB inspiral.
In the second panel we plot the integrated GW luminosity vs the frequency and see that $dE/df$ is larger for EsGB, such that there is a smaller change in frequency with a given emission of energy in that case. This effect tends to decelerate the EsGB inspiral.
The third panel displays the total energy emission in the time domain, demonstrating that a higher value is emitted (following alignment) in order to reach merger in the EsGB case. However, overall it still merges sooner (and at a higher frequency, meaning its frequency has needed to change more to reach merger).
Given that $df/dt \sim P ~ df/dE$, we therefore conclude that the observed acceleration of the EsGB merger (larger $df/dt$) must be the result of a greater overall emission of energy (larger $P$), which overcomes the conservative dynamics (smaller $df/dE$ compared to GR).
Another possible source of the difference is seen in the final panel, where we plot the change in irreducible mass of the BHs over time, showing that following their scalarization this quantity is larger than in GR. This may again contribute to the acceleration towards the merger, although the difference is small.
We find that the final ADM mass of the EsGB remnant is smaller by $\sim 1\%$, which is consistent with the larger radiation of energy.

If we assume that there is negligible horizon absorption of the scalar, these results suggest that modifications to the metric due to the presence of the scalar have changed the binding energy of orbits corresponding to given frequencies. More energy now needs to be emitted for the orbital frequency to increase by a given amount in the modified theory. Nevertheless, the radiation power must compensate for this, since we observe a faster merger rate overall. The exact cause of the different conservative dynamics is still to be fully understood, but is consistent with the EOB results in~\cite{Julie:2024fwy}.

\section{Notion of ADM mass in numerical simulations and PN calculations}
\label{sec:pn_mass}

We now briefly comment on how our ADM mass compares with the PN definition, since matching these is essential for comparing PN and NR waveforms and for setting consistent NR initial data. This is mostly based on the following references~\cite{Julie:2019sab,Julie:2022huo} to which we refer the reader for more details.

In GR PN theory, BHs are treated as point particles with constant masses and spins, and Einstein’s equations are solved as an expansion in $v/c$. In EsGB, however, it is convenient to 
``skeletonize'' the BHs in binary systems and replace the constant GR mass with a scalar-dependent function $\{m_i(\phi)\}_{i=A,B}$ determined by the value of the scalar field at the BH location~\cite{Eardley:1975fgi} (with A and B referring to each one of the two BHs). Focusing on a BH in isolation first, the value of  the function $m_i(\phi)$ was worked out explicitly in~\cite{Julie:2019sab} (cf. Eq. III.8). A skeletonized BH was found to be fully characterized by 
the asymptotic scalar field value $\phi_\infty(t)$ at spatial infinity and its constant irreducible mass,
\begin{align}
    \mu_i %
    = \sqrt{\frac{\mathcal{S}_W}{4\pi}},
\end{align}
where $\mathcal{S}_W$ is the Wald entropy (Appendix C of~\cite{Julie:2019sab}) which reduces in shift-symmetric EsGB to~\cite{Julie:2022huo}
\begin{align}
    \mathcal{S}_W = \frac{\mathcal{A}_H}{4} + 8\pi\lambda\,\phi_{\rm H}.
\end{align}
This entropy satisfies the first law of thermodynamics and is the sum of the Bekenstein term and a Gauss-Bonnet contribution. The key insight of Ref.~\cite{Julie:2019sab} was that when placing the BHs in a binary, the value of $\phi_{\infty}$ can no longer be
set to zero and will depend on a far-away companion. However, when doing PN calculations, we can assume that the BHs are initially very far apart, which allows us to solve the
Einstein equations not only as an expansion in $v/c$ but also around a constant background
scalar field value $\phi_0$, such that $\phi_{\infty} = \phi_0 + \delta \phi$, where $\delta \phi$ is at least $\mathcal{O}(1/r)$ and $r$ is orbital separation radius. In the case of shift-symmetric EsGB, where the field equations are invariant under shifts in the scalar field, the value of $\phi_0$, which is set by the binary's cosmological environment, can be set to zero without loss of generality. Finally, the masses of BHs that enter all PN expressions for the potential, fluxes and waveforms are the $m_i(\mu_i,\phi_0) \equiv m_i^0$ (see e.g Eq.(IV.6) in \cite{Julie:2019sab}). 

In our simulations we computed the Wald entropy of each BH and found that it changes by $\sim \{0.6,0.3\}\%$ for the smaller and bigger BH during scalarization, respectively, and remains nearly constant thereafter. Using these entropies, we evaluated the irreducible mass for each BH and obtained a total mass of $\mu_1+\mu_2$, which differs by $\sim 0.6\%$ from its initial value, as displayed in the last panel of Fig. \ref{fig:GWPower}. It is possible that this explains the higher frequency reached in EsGB, although further work is needed to properly understand the phenomenon. On the other hand, the change in the total ADM mass remains less than $0.1\%$, which we would not expect to have a significant impact on the results. We do not pursue mass-definition ambiguities further, although this remains a subtle issue even in GR (see~\cite{Sun:2025una}). 

%
\section{Discussion}
%
Studying the dynamical and non-linear regime of modified theories of gravity remains a major theoretical challenge, limiting our ability to place the most stringent constraints on these models. In recent years, several groups have independently solved the full equations of motion for a broad class of modified theories of gravity. Numerical codes are now capable of producing complete inspiral-merger-ringdown waveforms for use in injection campaigns, making it essential to validate results obtained through different numerical approaches. In this work, we used two distinct yet equally successful formulations of the Einstein equations: the modified generalized harmonic gauge and the modified CCZ4 formulation. We applied them to shift-symmetric EsGB gravity and showed that in this theory the late inspiral/merger phase evolution accelerates relative to GR, but to a much smaller extent than expected from PN predictions.

We focused on a roughly equal-mass case for which the dipolar emission is suppressed, but it is possible that more unequal-mass cases also show reduced dephasing at late stages where the tensor radiation is dominant. Longer waveforms beginning at larger separations would better characterize the transition from the PN regime to the fully non-linear case studied here, in which conservative dynamics appear to play a significant role. If the transition were to happen within the observational band of GW detectors, it may appear in inspiral–merger–ringdown consistency tests~\cite{LIGOScientific:2019fpa,Abbott:2020jks,LIGOScientific:2020tif,LIGOScientific:2021sio,LIGOScientific:2025obp,LIGOScientific:2025rid}, since it would be difficult to mimic with a simple change to the inferred masses. It should also be different from the imprints of environments such as dark matter \cite{Bamber:2022pbs,Aurrekoetxea:2023jwk,Aurrekoetxea:2024cqd,Machet:2025vzt,Roy:2025qaa}, which tend to strongly accelerate the merger relative to GR due to increased dissipative effects (although see recent work in \cite{Cheng:2025wac}).

We emphasize that the significance of this result is not limited to this particular theory, which as we have already noted is quite constrained by observations. It demonstrates that the effect of a modified metric and its impact on the binary dynamics can play a significant role, particularly in the strong field regime prior to merger. Distinctive effects have recently been observed in the context of oppositely charged black-hole binaries in scalar–Gauss–Bonnet theories near the scalarization threshold \cite{Lara:2025kzj}, where non-linear dynamics were found to increase eccentricity in the late inspiral, contrary to expectations that extra degrees of freedom enhance circularization. Such results underscore that non-linear strong-field dynamics can qualitatively overturn standard heuristics derived from weak-field or perturbative reasoning. They highlight the crucial role of full numerical solutions in validating modified gravity waveforms.

The next step is to define the region of the parameter space where the transition from faster to slower inspiral happens. This will depend on the mass ratio, coupling values and range of frequencies probed. Longer waveforms in the shift-symmetric theory were recently completed using another formulation and implementation of the Einstein equations~\cite{Lara:2024rwa,Lara:2026xcs,Lara:2026dwh}, which show comparably small dephasing, and scattering studies have also been performed in a precision regime~\cite{Swain:2026dfr}. Injections can be used to determine how degenerate the signal is with other effects like spin-induced precession, and whether it can trigger the current tests of GR used by the LVK Collaboration, or results in a distinctive shift in the intrinsic binary parameters. 
There is also the exciting prospect of studying a wider range of theories, including parity violating and pure tensor theories, due to recent groundbreaking developments in well-posed formulations~\cite{Figueras:2024bba,Figueras:2020ofh}. The ultimate goal of this programme of work is to catalogue these kinds of smoking-gun effects across a range of theories, in order to discover or constrain EFT modifications using future observations. Our findings indicate that systematic cross-correlation between independent codes and complementary approximations are essential for revealing subtle phenomena that may otherwise remain undetected. 

\

\acknowledgements
We acknowledge helpful conversations with A.~Buonanno,S.~Brady, P.~Figueras, S.~Gasparotto, T.~Jain, F.L.~Juli\'e, \'A.~Kov\'acs, G.~Lara, L.~Lehner, L.~Pompili, S.~Swain and J.~Zosso. We thank D.~Doneva, W.~East, P.-J.~Nee, H.~Pfeiffer, T.~Sotiriou and N.~Yunes for providing useful comments on the manuscript. 
LAS has been partly funded by Interuniversitaire Bijzonder Onderzoeksfonds (IBOF)/21/084.
KC acknowledges support from the Simons Foundation International and the Simons Foundation through Simons Foundation grant SFI-MPS-BH-00012593-13, a UKRI Ernest Rutherford Fellowship (grant number ST/V003240/1) and an STFC Research Grant ST/X000931/1 (Astronomy at Queen Mary 2023-26).
KC and LAS thank the GRTL collaboration (\url{www.grtlcollaboration.org}) for their support and code development work.
Some of the computations were performed on the Urania HPC system at the Max Planck Computing and Data Facility.
This work used the DiRAC Memory Intensive service Cosma8 at Durham University, managed by the Institute for Computational Cosmology on behalf of the STFC DiRAC HPC Facility (\url{www.dirac.ac.uk}). The DiRAC service at Durham was funded by BEIS, UKRI and STFC capital funding, Durham University and STFC operations grants. DiRAC is part of the UKRI Digital Research Infrastructure. We also used the resources provided by the VSC (Flemish Supercomputer Center), funded by the Research Foundation Flanders (FWO) and the Flemish Government.
We acknowledge as well the EuroHPC Joint Undertaking for awarding this project access to the EuroHPC supercomputer LUMI-C, hosted by CSC (Finland) and the LUMI consortium through a EuroHPC
Extreme Scale Access call, under project EHPC-EXT-2025E02-070.

\vspace{-0.5cm}

\bibliography{main.bib}

\clearpage

\appendix


\section{Method of alignment}

Here we explain in detail the method used for aligning the waveforms.
The waveforms are aligned as was done in Ref.~\cite{Julie:2024fwy} by first minimizing the integral
\begin{equation}\label{eq:align_freq_1}
        \mathcal{I}_f(\delta t) = \int_{t_i}^{t_f} dt \left| f_{\rm{GR}}
        - f_{\rm{GB}} (t + \delta t)  \right |^2,
\end{equation}
then
\begin{equation}\label{eq:align_freq_2}
        \mathcal{I}_{\Phi}(\delta \Phi) = \int_{t_i}^{t_f} dt \left| \Phi_{\rm GR}
        - \Phi_{\rm GB} (t + \delta t) + \delta \Phi \right |^2,
\end{equation}
with $\delta \Phi$ and $\delta t$ being relative phase and time shifts, $\Phi_{\rm GR}$
and $\Phi_{\rm GB}$ denote the phases of the GR and EsGB waveforms. The alignment is done in a time window
$[t_i,t_f ]$ that corresponds to a low-frequency interval ($0.0074\lesssim f_0\,M\lesssim 0.0083$).
The lower bound of the window should be chosen as early as possible, but late enough for junk radiation to dissipate and BHs to scalarize. We do not find that our results depend sensitively on the starting time of alignment. The upper bound is arbitrary, but should be large enough to average out numerical noise and residual eccentricity.
We show in Fig.~\ref{fig:window_dependence} that the results do not sensitively depend on the upper bound we choose.

\begin{figure}[t]
\includegraphics[width=\columnwidth]{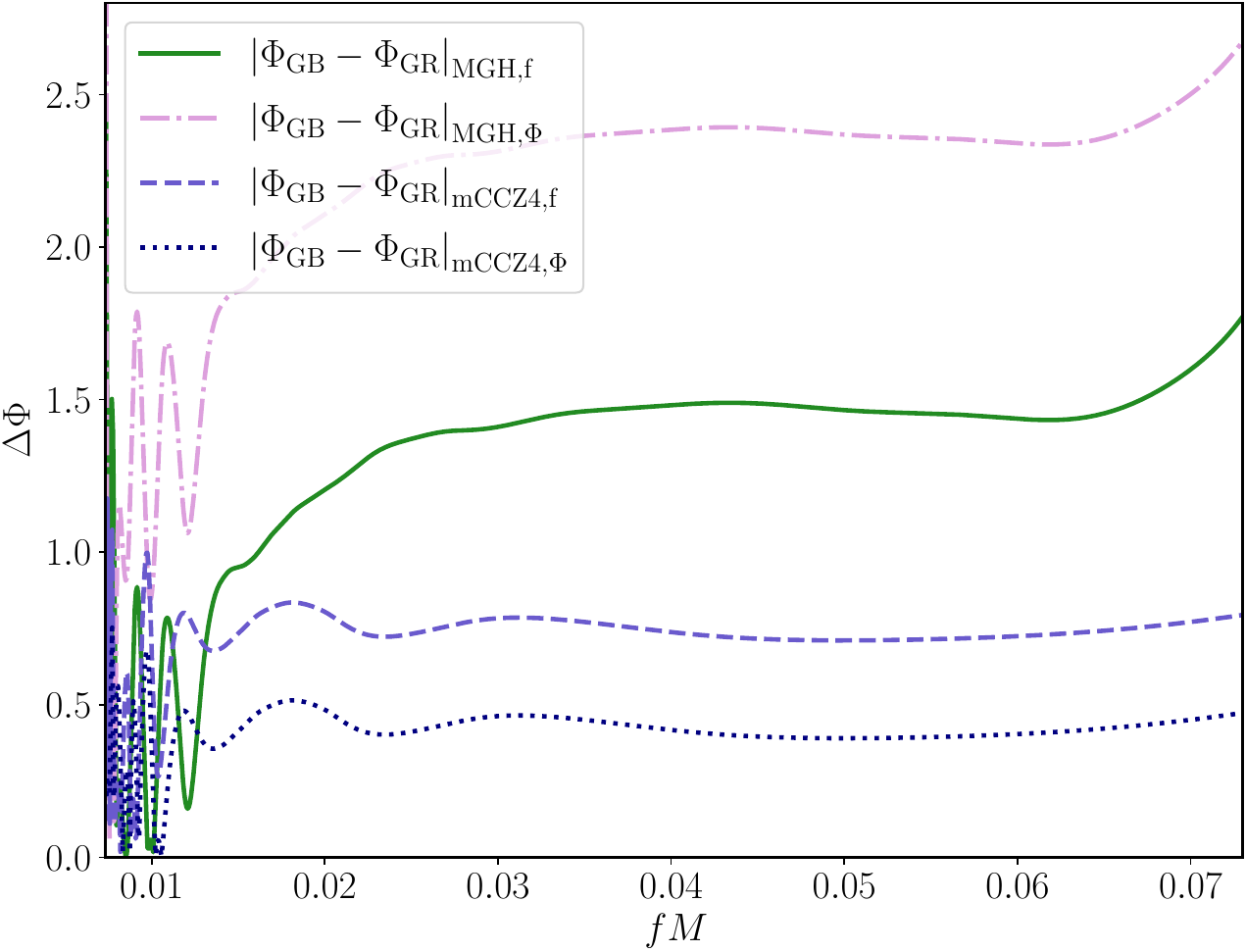}
\caption{Sensitivity of results on the way we perform alignment. Dephasing between GR and EsGB waveforms for each formulation after aligning the waveforms. The subscript $f$ refers to alignment according to \eqref{eq:align_freq_1}-\eqref{eq:align_freq_2}, while subscript $\Phi$ to alignment performed using  \eqref{eq:align_phase}. 
}\label{fig:convergence_alignment}
\end{figure}

\begin{figure}[t]
\includegraphics[width=\columnwidth]{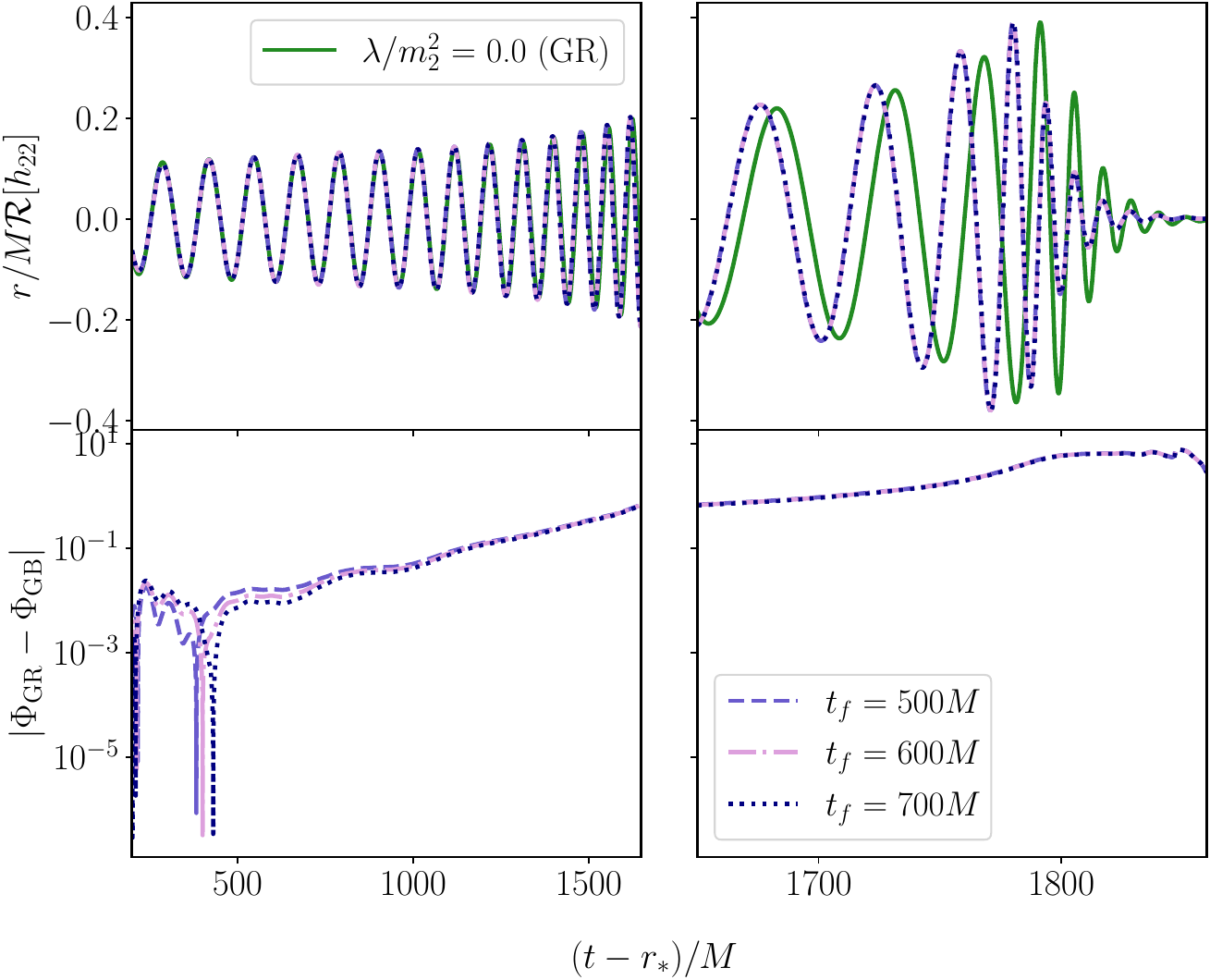}
\caption{Sensitivity of results on the upper bound for alignment. Top: Strain as a function of retarded time for GR (solid green) and EsGB after aligning the waveforms according to \eqref{eq:align_freq_1}-\eqref{eq:align_freq_2} with $t_i=200M$ but varying $t_f=\{500,600,700\}M$. Bottom: Dephasing between GR and EsGB waveforms for the different window lengths of the top panel. We find that the alignment does not sensitively depend on the time window we choose.}\label{fig:window_dependence}
\end{figure}

We also compared to minimizing the integral
\begin{equation}\label{eq:align_phase}
        \mathcal{I}(\delta t,\delta \Phi) = \int_{t_i}^{t_f} dt \left| \Phi_{\rm GR}
        - \Phi_{\rm{GB}} (t + \delta t) + \delta \Phi \right |^2,
\end{equation}
as was done in Ref.~\cite{Taracchini:2012ig}, and show in Fig.~\ref{fig:convergence_alignment} that the results for both codes do not sensitively depend on the way we do the alignment.

\begin{figure}[]
\includegraphics[width=\columnwidth]{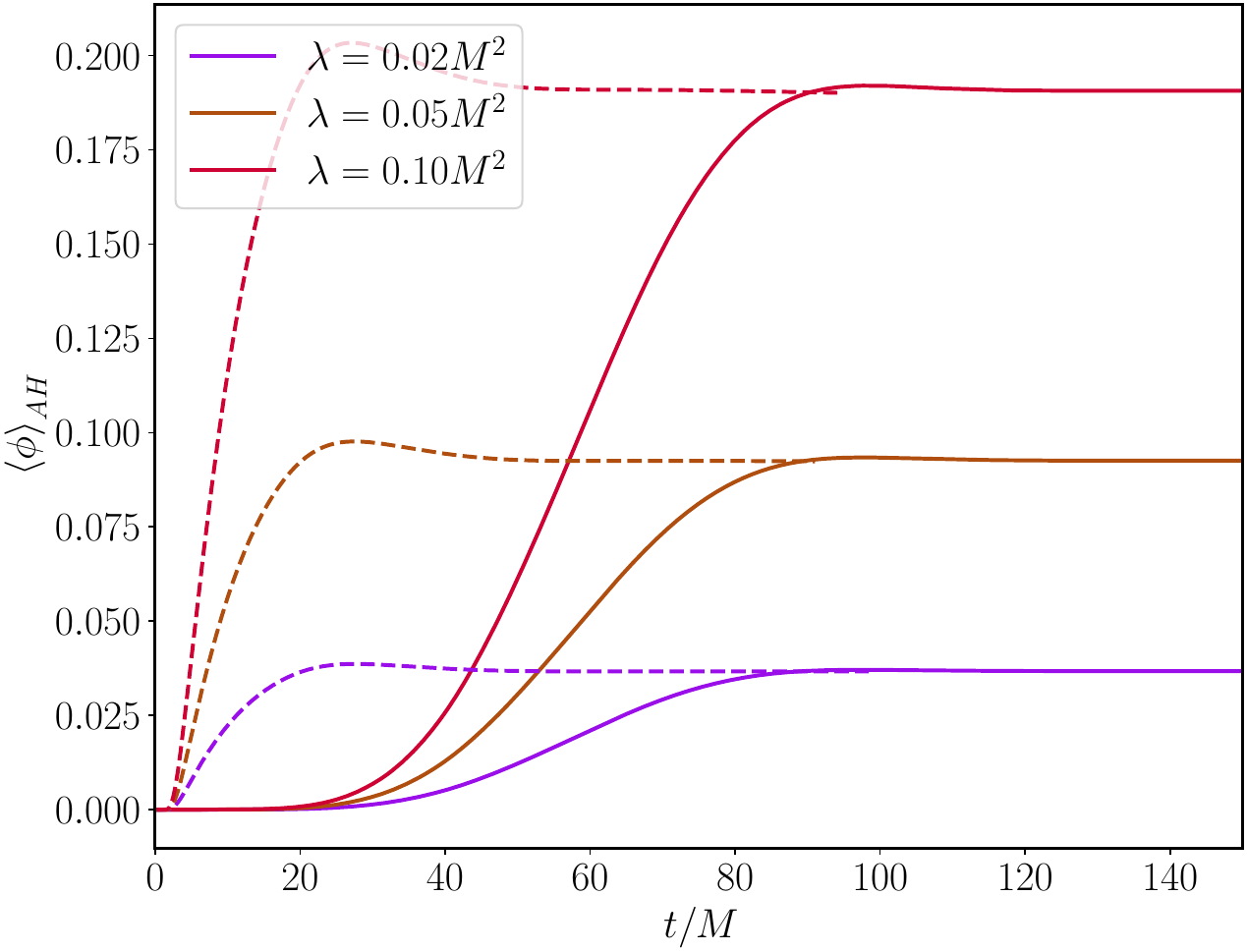}
\caption{Scalarization of isolated non-spinning BHs: Evolution in time of the average value of the scalar field at the Apparent Horizon for different values of the coupling constant, with dotted and solid lines corresponding to the mCCZ4 and MGH codes, respectively. Note that the scalarization takes longer in the MGH code because the coupling is slowly ramped up on a timescale of $100M$ in an analogous way as done for the BBH in this work.}\label{fig:single_bh}
\end{figure}


\begin{figure}[t]
\includegraphics[width=\columnwidth]{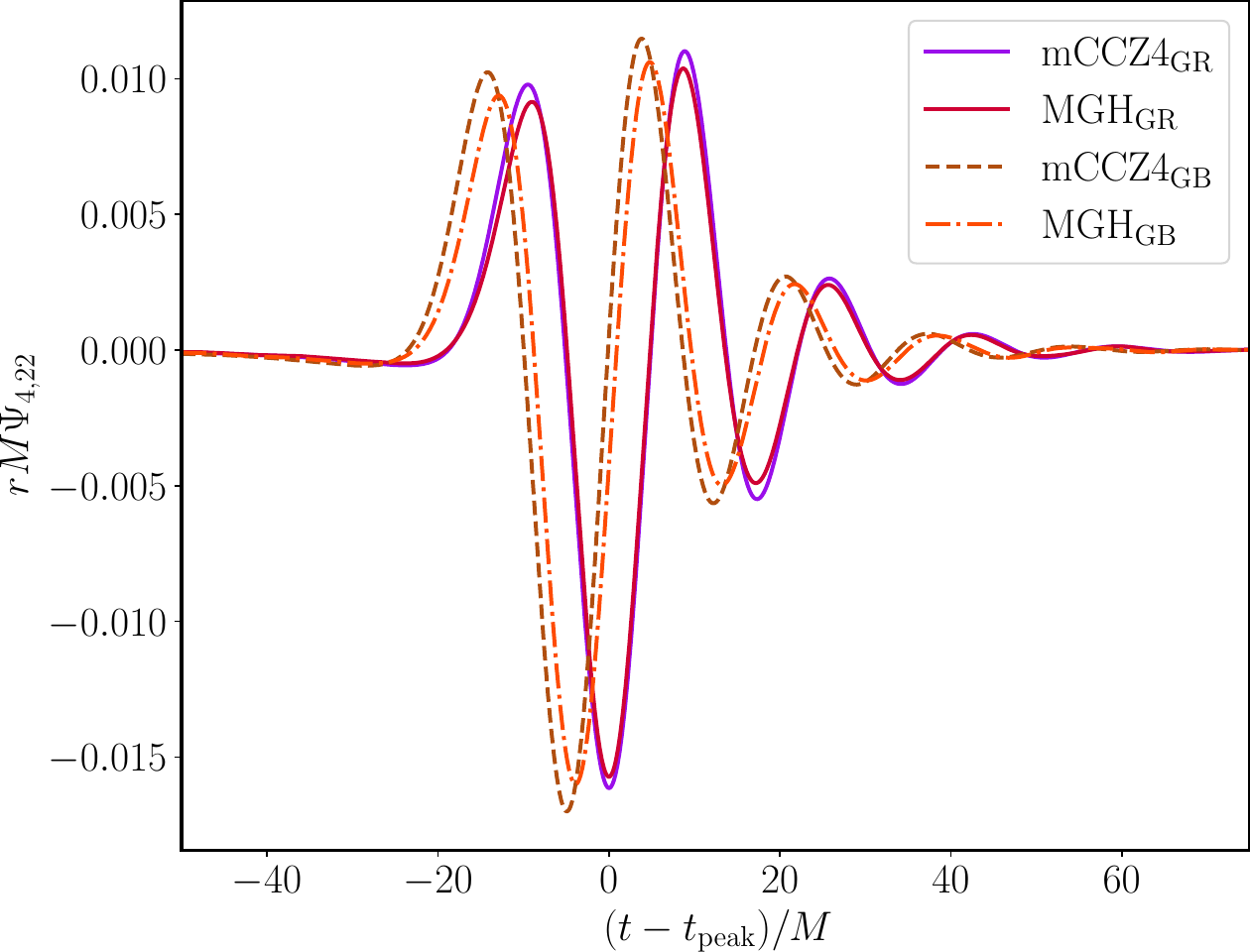}
\caption{Equal mass head-on mergers in EsGB and GR: Evolution in time of the (2,2) mode of the Weyl scalar extracted at spatial infinity. The differences between the two codes are compatible with truncation errors.}\label{fig:headon}
\end{figure}

\section{Code comparison}\label{sec:compare}

In this section we provide a detailed description of the two independent numerical implementations used in this work, the MGH and mCCZ4 formulations, for evolving the full, non-perturbative, shift-symmetric EsGB equations.
In both formulations, one introduces two auxiliary
Lorentzian metrics $\tilde{g}^{mn}$ and $\hat{g}^{mn}$
that obey certain causality conditions. Adopting a gauge in these formulations amounts to choosing the functional form of the auxiliary metrics,
as well as the functional form of the source function $H^c$.
There is significant freedom in choosing the auxiliary metrics as functions of the physical metric $g_{ab}$, but
as in Refs.~\cite{East:2020hgw,AresteSalo:2022hua} we choose
 $\tilde{g}^{ab} = g^{ab} - \tilde{A} n^{a}n^{b}$ and
$ \hat{g}^{ab} = g^{ab} - \hat{A} n^{a}n^{b}$, where $n^{a}$ is the (timelike) unit vector orthogonal to the spacelike hypersurfaces we evolve on, and $\tilde{A}$ and $\hat{A}$ are constants set to 0.2 and 0.4, respectively.
However, the two approaches differ in the choice of source function.
The MGH formulation uses the modified version of the damped harmonic gauge, sometimes fixed in time to maintain Kerr-Schild like coordinates \cite{Lindblom:2009tu,Choptuik:2009ww}. The mCCZ4 scheme, on the other hand, employs a generalized moving-puncture gauge based on 1+log slicing and Gamma-driver conditions.
Both methods discretize the evolution equations with fourth-order Runge–Kutta time integration and fourth finite-difference stencils, supplemented by Kreiss–Oliger dissipation~\cite{1972Tell...24..199K} and constraint damping.
The MGH formulation directly evolves the covariant system, inverting the resulting linear equations at each grid point using Gaussian elimination, and employs compactified coordinates, as detailed in Ref.~\cite{Pretorius:2004jg}, so that physical boundary conditions
(namely that the metric is flat and the scalar field vanishes)
can be placed at spatial infinity. In contrast, the mCCZ4 formulation uses a finite Cartesian grid with standard Sommerfeld boundaries and a 3+1 conformal decomposition of the equations.
Both utilize Berger-Oliger~\cite{1984JCoPh..53..484B} style adaptive mesh refinement (AMR), implemented via the PAMR~\cite{Pretorius:2005ua,PAMR_online} and Chombo~\cite{Adams:2015kgr} libraries, respectively with third-order temporal interpolation for the AMR boundaries.
Treatment of BHs also differs: the MGH approach relies on dynamic excision of regions within apparent horizons, with second-order extrapolation to repopulate grid points as horizons evolve, while the mCCZ4 scheme uses the moving-puncture technique to avoid explicit excision and smoothly suppresses the higher-derivative terms inside horizons. Further implementation details can be found in the respective references.

Here we show additional results, complementary to the quasi-circular BBH studied in the main body of the manuscript, which helped us validate our codes. 

We first focus on the scalarization of isolated non-spinning BHs for different coupling parameters $\lambda$, ranging from $\lambda/M^2 = 0.02$ to $0.1$. Figure~\ref{fig:single_bh} shows the evolution in time of the average value of the scalar field on the BH apparent horizons, $\langle\phi\rangle_{\rm AH}$, for three different values of the coupling constant, with dotted and solid lines corresponding to the mCCZ4 and MGH codes, respectively, with excellent agreement in the values attained after scalarization.

We then study equal mass head-on collisions with an initial separation of $50M$ for $\lambda/M^2=0.025$, and compare the gravitational radiation to GR. Figure~\ref{fig:headon} displays the (2,2) mode of the Weyl scalar for both GR and EsGB, showing very good agreement between the two codes. We have shifted the waveforms in time so that both MGH and mCCZ4 codes merge at $t=0$ in GR and we have kept the same negative dephasing between GR and EsGB present in the raw data. Given that head-on mergers have no phase evolution, there is no notion of frequency and so there is no obvious way to align GR and non-GR waveforms. Since the initial transients induce a change of the velocities with which the punctures are moving towards each other, the GR and EsGB waveforms may represent different physical systems. However, this does not change the main message of Figure~\ref{fig:headon}, which is the agreement of our codes for both GR and beyond GR head-on mergers. 

\section{Comparison to first version of paper and convergence test}\label{sec:prev_version}
In a previous version of this paper it was stated that the EsGB binary was merging later than the GR system, but on more careful study we now find the opposite, although we still find that the amount of dephasing is significantly below that predicted by PN. 
In this section we compare our results with those of the first version of this paper. The results for the MGH formulation were not updated, whereas those for the mCCZ4 formulation are. We show below that this apparent discrepancy is accounted for by the truncation error of the earlier simulations: in both formulations the older runs are consistent with either sign of the dephasing once the numerical error is properly estimated, and the new, better-resolved mCCZ4 runs give a dephasing that is consistently negative and compatible with the older results within their error bars.

\begin{figure}[t]
\includegraphics[width=\columnwidth]{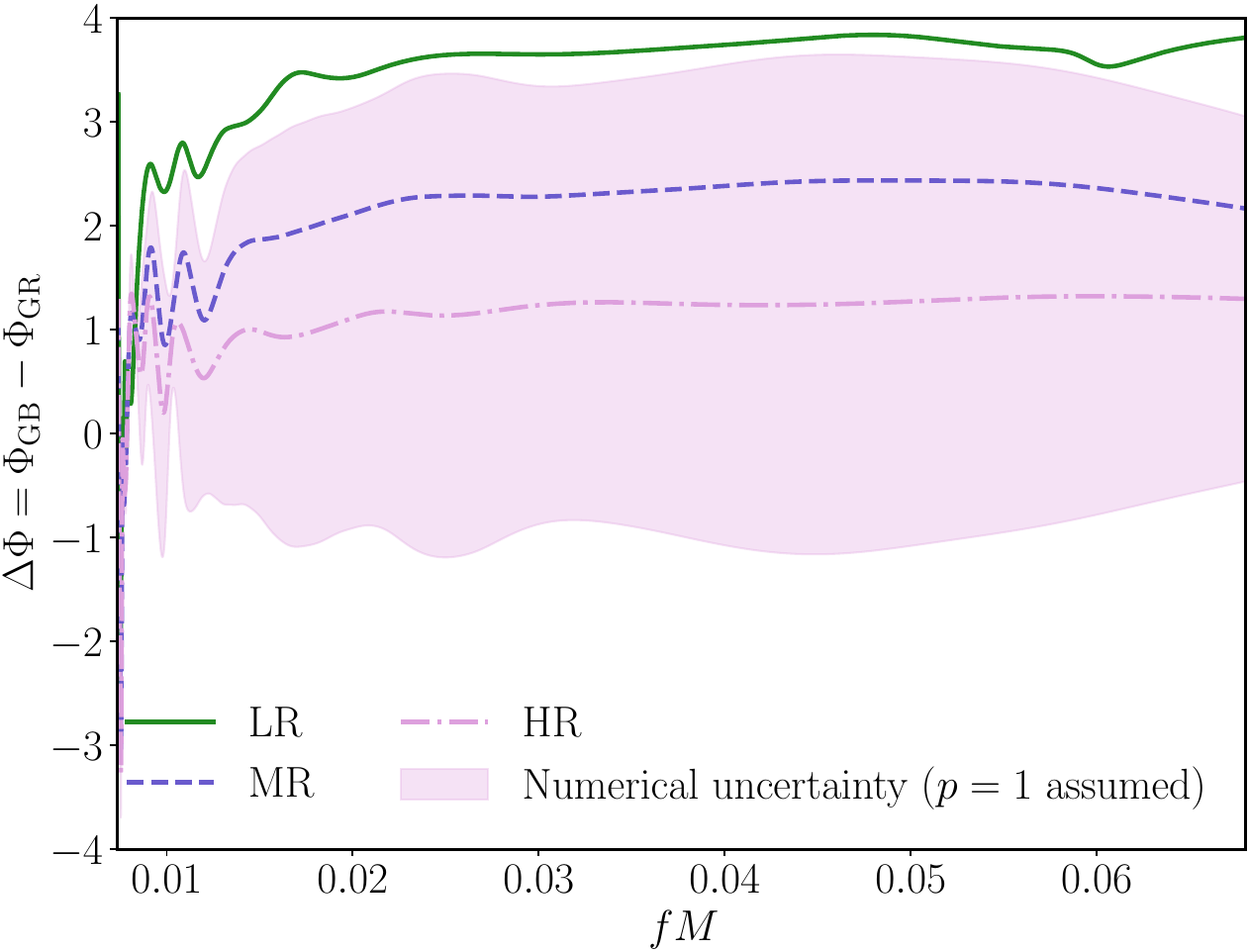}
\includegraphics[width=\columnwidth]{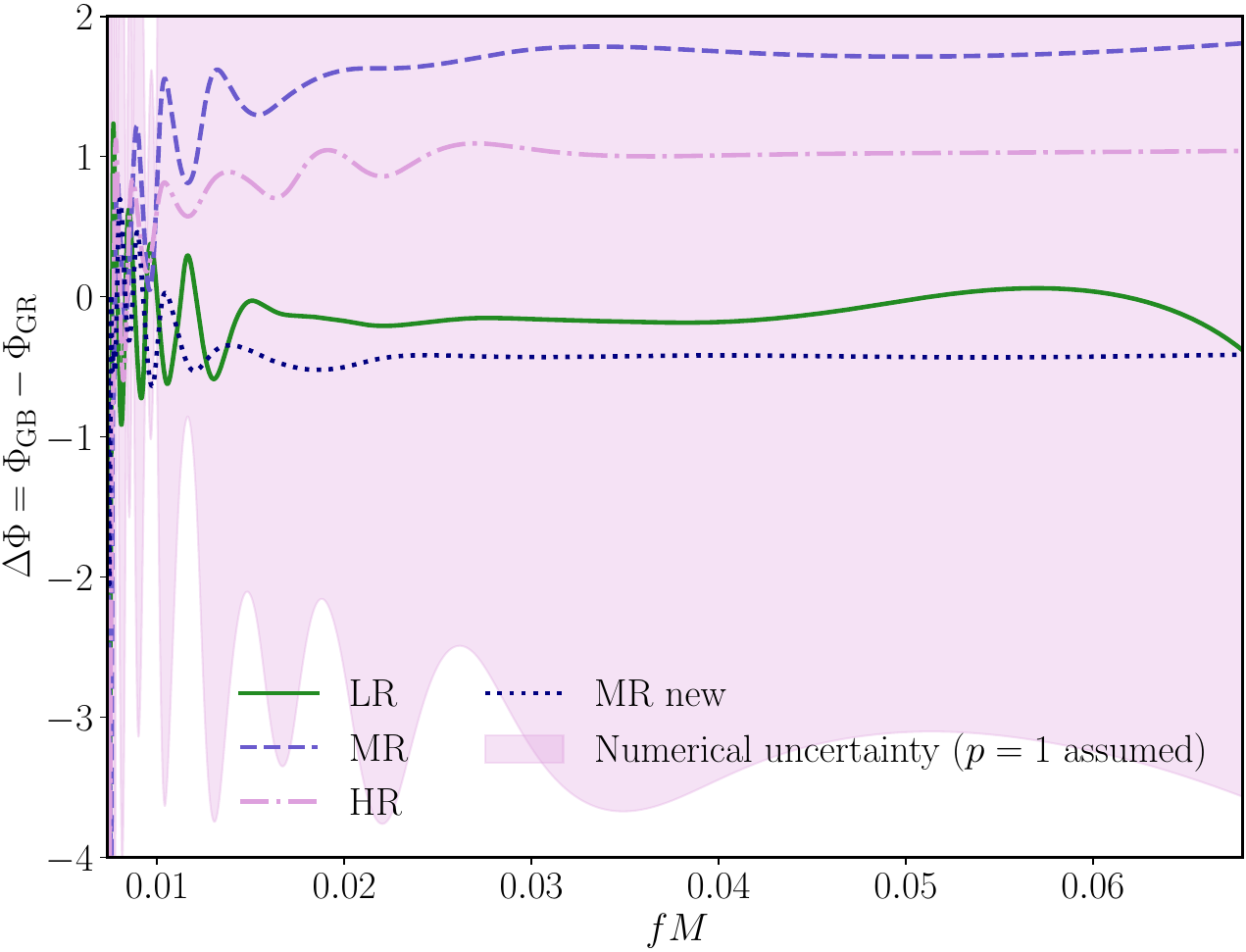}
        \caption{Dephasing as a function of frequency for the GW150914 binary black hole merger at three resolutions, for the MGH (top) and mCCZ4 (bottom) formulations. The higher- and lower-resolution runs use grid spacings $2/3\times$ ($6/7\times$) and $3/2\times$ ($6/5\times$) the reference value for the MGH (mCCZ4) formulation. The resolution-to-resolution change in the dephasing does not follow a clean power law at these resolutions, so rather than a Richardson extrapolation we show a shaded uncertainty band around the highest-resolution curve, obtained by rescaling the difference between the two highest resolutions assuming a conservative, first-order ($p=1$) convergence rate.
For the mCCZ4 formulation we additionally show the default-resolution dephasing of the new runs, which lies within this conservative uncertainty band.}\label{fig:dephasing_freq_rich}
\end{figure}

We first consider the numerical uncertainty of the MGH runs. These simulations use seven
levels of adaptive mesh refinement with a refinement ratio of 2:1 and a grid spacing of
$dx=0.016M$ on the finest level. In Fig.~\ref{fig:dephasing_freq_rich} we plot the
dephasing as a function of frequency at the default resolution and at two additional
resolutions, with grid spacings $3/2$ and $2/3$ $\times$ the reference value. 
Although the phase evolutions of the gravitational GR and EsGB cases, taken individually, show consistent second-order convergence,
the change in the phase \emph{difference} (i.e. the difference between the phases after alignment) 
between resolutions is large compared to the signal itself and
does not follow a clean power law, so rather than a Richardson extrapolation we show the
raw dephasing at all three resolutions together with a conservative numerical-uncertainty
band around the highest-resolution result. This band is obtained by rescaling the
difference between the two highest resolutions assuming a conservative first-order
($p=1$) convergence rate; although the central (highest-resolution) value corresponds to
a delayed merger in EsGB, the uncertainty band is equally consistent with the opposite
trend.

We repeat this exercise for the mCCZ4 formulation, for which we have two sets of runs.
The first set, presented in the first version of the paper, uses ten refinement
levels with a grid spacing of $dx=0.008M$ on the finest level
and two additional resolutions at $6/5$ and $6/7$ $\times$ the reference
value. As shown in Fig.~\ref{fig:dephasing_freq_rich}, the behaviour mirrors the MGH case:
the change between resolutions is again too large to support a clean convergence order,
so we show the same conservative, assumed-first-order uncertainty band. The
highest-resolution result is positive but, within this uncertainty band, still consistent
with the opposite trend.
For the mCCZ4 formulation we also performed a new set of simulations, which provide
the main results of the current version of the article. These runs differ from the
previous ones in several respects, the most important being the larger size of the finest level
around each one of the punctures, which we set here to be $m_i+1.2M$, instead of $1.5m_i$.
This significantly reduces the numerical noise, which allows us to use a smaller reference resolution, with a grid spacing of $dx=0.0125M$
on the finest level, while obtaining more accurate results\footnote{We note that the set-up used for the new runs was motivated by the testing carried out in the scattering runs of ~\cite{Swain:2026dfr}.}. 
Here the dephasing is consistently negative across all three resolutions, and the resolution-to-resolution
change is significantly smaller than for the earlier runs. The reference-resolution
result, also shown in Fig.~\ref{fig:dephasing_freq_rich}, lies within the (conservative,
assumed first-order) uncertainty band of the earlier mCCZ4 runs.

\begin{figure}[t]
\includegraphics[width=\columnwidth]{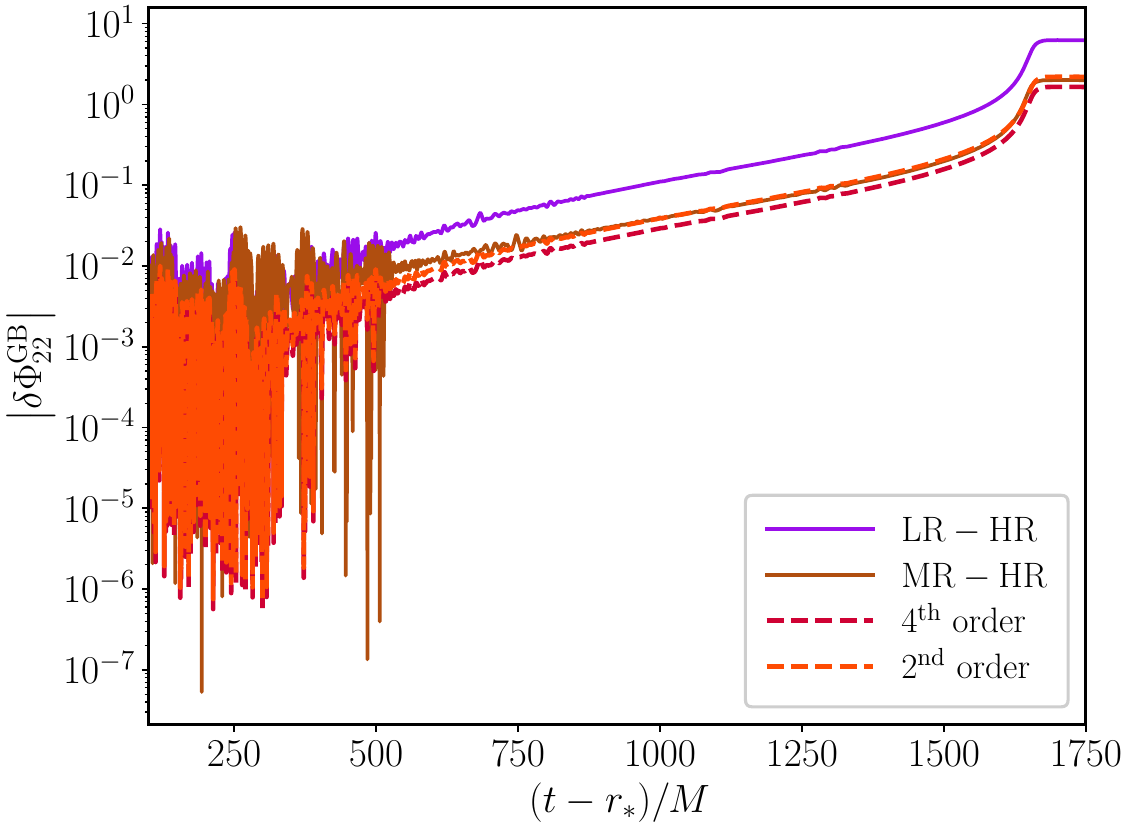}
\includegraphics[width=\columnwidth]{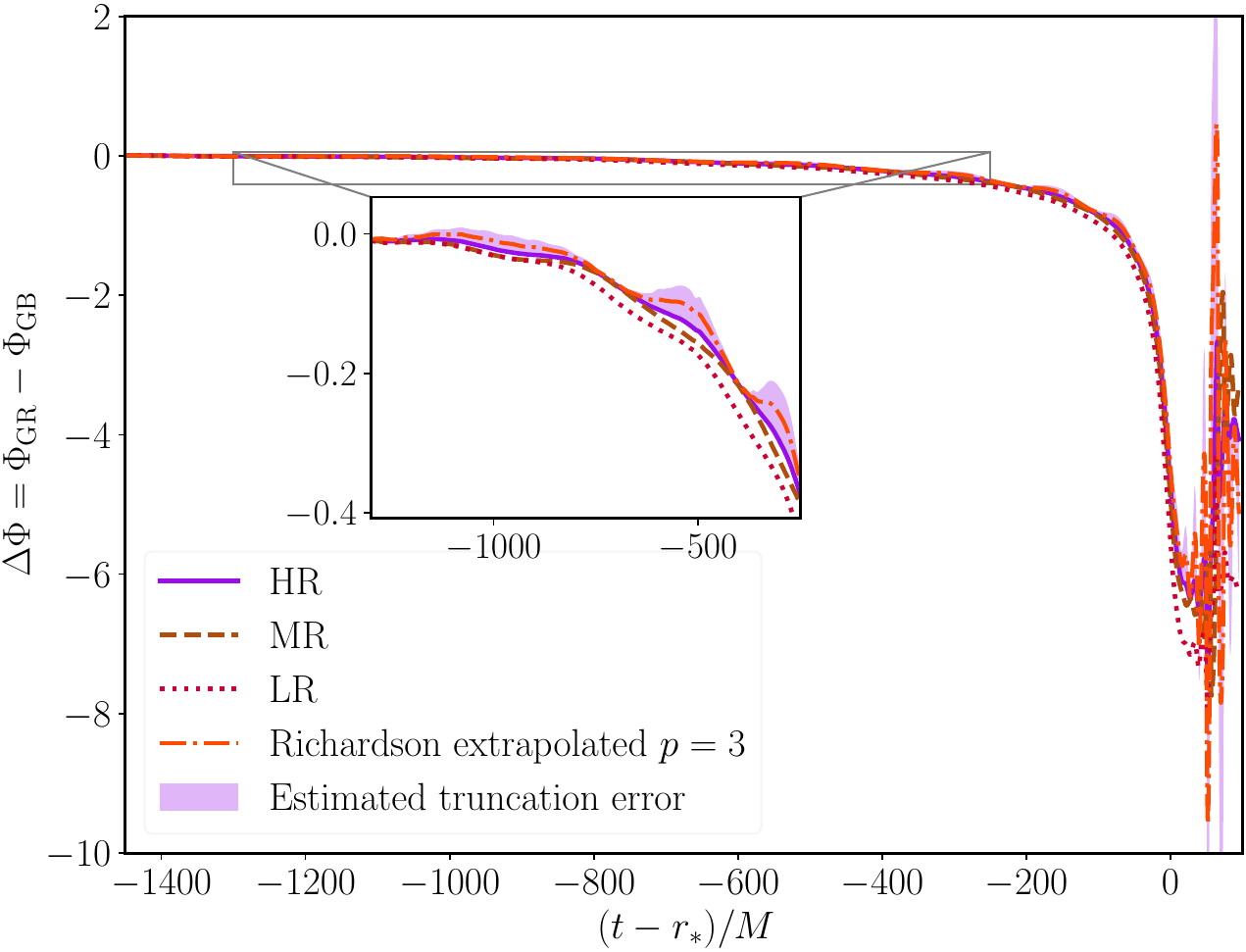}
        \caption{Convergence plots for the GW150914 binary black hole merger at three resolutions for the mCCZ4 formulation presented in main text. The higher- and lower-resolution runs use grid spacings $1.2\times$ and $1/1.25\times$ the default value. Left: Difference between low and high, and medium and high resolutions as a function of time for GB simulation. We find roughly third order convergence.
        Right: Convergence of phase of GB simulations as a function of time. The convergence is close to
$p=3$. We also show the Richardson-extrapolated dephasing and an estimate of the
truncation error.}\label{fig:convergence_new}
\end{figure}

In Fig.~\ref{fig:convergence_new} we test the convergence of the new mCCZ4
runs, complementing the frequency-domain comparison above. In contrast to the older
mCCZ4 and MGH runs discussed above, for which no consistent convergence order could be
established and the resolution-to-resolution change in the dephasing was comparable to,
or larger than, the dephasing signal itself, here the difference between the medium- and
high-resolution dephasing is only about $9\%$ of the signal. The left panel shows the
low$-$high and medium$-$high differences of the EsGB phase as a function of time; the two
curves track each other over close to two orders of magnitude once rescaled assuming
third-order convergence. This justifies computing a genuine Richardson-extrapolated
dephasing and an associated truncation-error estimate for the new runs, shown in the
right panel, in contrast to the earlier runs of both formulations, for which we instead
report the conservative uncertainty band of
Fig.~\ref{fig:dephasing_freq_rich}.

\section{Comparison to previous works}\label{sec:prev_results}

Finally, we compare the effects presented in this paper with our earlier works, in particular \cite{Corman:2022xqg} and \cite{AresteSalo:2025sxc}, and account for the differences observed - in these previous works, in general we found a larger amount of dephasing than is reported here.
In the MGH simulations of \cite{Corman:2022xqg}, no eccentricity–reduction procedure was applied, and the EsGB waveforms were slightly more eccentric than their GR counterparts due to initial transients caused by BH scalarization. The available waveforms were also too short to allow an alignment over the longer time interval used here—an interval that is crucial for averaging out numerical noise and residual eccentricity. In the mCCZ4 runs of \cite{AresteSalo:2025sxc}, the waveform alignment between the GR and EsGB waveforms was also not done using the robust method used here, and the value of the coupling studied was lower. In the unequal mass cases the presence of a larger amount of dipole radiation may also have reduced the effect, and although the eccentricity was similar between the GR and EsGB cases, the fact that the orbits were non-circular will have changed the dynamics and reduced the ability to robustly align them. 
By contrast, the waveforms employed in the present study are substantially longer, have undergone eccentricity reduction, and therefore permit a robust alignment procedure. This improved setup was essential for enabling the rigorous code comparison and quantification of the dephasing.

Note that hints of the reduction in dephasing have already appeared in fully non-linear, quasi-circular, and eccentricity-reduced simulations of black hole–neutron star mergers in Einstein–scalar–Gauss–Bonnet gravity using the MGH formulation~\cite{Corman:2024vlk}. In that work, even though the neutron star does not carry scalar charge and scalar radiation is therefore expected to be non-negligible, it was found that the additional dephasing beyond the inspiral stage is negligible (see Fig. 3 of Ref.~\cite{Corman:2024vlk}).

\end{document}